\documentclass[12pt]{article}
\usepackage{latexsym,graphicx,epsfig,psfrag,here}
\newcommand{\beq}{\begin{equation}} 
\newcommand{\eeq}{\end{equation}} 
\newcommand{\beqa}{\begin{eqnarray}} 
\newcommand{\eeqa}{\end{eqnarray}} 
\newcommand{\beqan}{\begin{eqnarray*}} 
\newcommand{\eeqan}{\end{eqnarray*}} 
\newcommand{\ba}{\begin{array}} 
\newcommand{\ea}{\end{array}} 
\newcommand{\no}{\nonumber} 
\newcommand{\grts}{\stackrel{>}{_\sim}}

\newcommand{\ve}{\varepsilon}

\newcommand{\nn}{\nonumber \\}

\newcommand{\bea}{\begin{eqnarray}} 
\newcommand{\eea}{\end{eqnarray}}

\newcommand{\IM}{\mbox{\rm Im}} 
\newcommand{\hepph}[1]{{\tt hep-ph/#1}}

\normalsize 
\begin{document} 
\begin{titlepage} 
\begin{flushright} 
UWThPh-2000-54\\ 
LNF-01/002(P)\\ 
Dec. 2000\\ 
\end{flushright} 
\vspace{2.5cm} 
\begin{center} 
{\Large \bf The CP-Violating Asymmetry in $K_L \to \pi^+\pi^- e^+
e^-$~*} \\[40pt] 
G. Ecker$^{1}$ and H. Pichl$^{1,2}$  
 
\vspace{1cm} 
${}^{1)}$ Institut f\"ur Theoretische Physik, Universit\"at 
Wien\\ Boltzmanngasse 5, A-1090 Vienna, Austria \\[10pt] 
 
${}^{2)}$ INFN, Laboratori Nazionali di Frascati, P.O. Box 13, I-00044
Frascati, Italy \\[10pt] 

\vfill 
{\bf Abstract} \\ 
\end{center} 
\noindent 
We update the theoretical analysis of the CP-violating asymmetry
in the decay $K_L \to \pi^+\pi^- e^+ e^-$, relying on chiral
perturbation theory and on the most recent phenomenological
information. With the experimentally determined magnetic amplitude 
and branching ratio as input, the asymmetry can be calculated with
good accuracy. The theoretical interpretation of the sign of the
asymmetry is discussed.
 
\vfill 
\noindent *~Work supported in part by TMR, EC-Contract  
No. ERBFMRX-CT980169 (EURODA$\Phi$NE).
 
\end{titlepage} 
\addtocounter{page}{1} 
\paragraph{1.}
A large CP-violating asymmetry in the decay $K_L \to \pi^+\pi^- e^+ e^-$
was originally predicted by Sehgal and Wanninger \cite{SW92}. The
effect is almost entirely due to indirect CP violation in
$K^0-\bar{K^0}$ mixing.
The predicted asymmetry in the angle between the $\pi^+\pi^-$ and the 
$e^+e^-$ planes has been confirmed experimentally \cite{ktev00,na4800}. 
The purpose of this note is to update the theoretical analysis of
the asymmetry \cite{SW92,HS93,ESW95,ESWW96,Sav99,Seh99} using 
the most recent phenomenological input available and employing the
methods of chiral perturbation theory. The latter will mainly be
invoked to estimate higher-order corrections  
but also to interpret the observed sign of the asymmetry.

\paragraph{2.} 
The amplitude for the decay $K_L(p) \to \pi^+(p_1)\pi^-(p_2) e^+(k_+)
e^-(k_-)$ is expressed in terms of three invariant form factors
$E_1, E_2$ (electric) and $M$ (magnetic):
\begin{eqnarray} 
A(K_L \to \pi^+\pi^- e^+ e^-) &=&
\frac{e}{q^2}\overline{u}(k_-)\gamma^\mu v(k_+)~V_\mu \nn  
V_\mu &=& i E_1 p_{1\mu} + i E_2 p_{2\mu} + M
\ve_{\mu\nu\rho\sigma}p_1^\nu p_2^\rho q^\sigma \label{ff}\\
q &=& k_+ + k_- ~. \no
\end{eqnarray} 
We use the set of variables originally introduced by Cabibbo
and Maksymovicz \cite{CM65} for $K_{e4}$ decays. With this choice,
the form factors $E_1,E_2,M$ depend on $s_\pi$ (invariant mass
squared of the two pions), $q^2$ (invariant mass squared of the
leptons) and $\theta_\pi$ (angle of the $\pi^+$ in the
$\pi^+\pi^-$ center-of-mass system with respect to the dipion line 
of flight in the kaon rest frame). The two remaining Dalitz variables
are $\theta_l$, the corresponding angle for the positron, and $\Phi$,
the angle between the dipion and dilepton planes in the kaon rest
frame.

After integration over four of these variables, the differential decay
rate with respect to $\Phi$ assumes the general form\footnote{We use
the conventions of Ref.~\protect\cite{BCEG95}, in particular 
$\ve_{0123}=1$ and the metric $(+---)$.} 
\cite{PT68,BCEG95}
\begin{equation} 
\displaystyle\frac{d\Gamma}{d\Phi} = I_1 + I_2 \cos\Phi + I_3 \sin\Phi
+ I_4 \cos2\Phi + I_5 \sin2\Phi ~.
\end{equation} 
Under a CP transformation 
\begin{eqnarray} 
\cos\Phi \longrightarrow \cos\Phi \nn
\sin\Phi \longrightarrow - \sin\Phi
\end{eqnarray} 
so that non-zero $I_3$, $I_5$ signify CP violation. It turns out
that $I_3$ is very small in the standard model, being sensitive to direct
CP violation only \cite{HS93}. The quantity of interest here is 
$I_5$ that is almost exclusively due to indirect CP violation
\cite{HS93}. A convenient measure of this term is the asymmetry
\begin{equation} 
{\cal A}_{CP} = \langle {\rm sgn}(\sin\Phi \cos\Phi) \rangle = 
\displaystyle\frac{4 I_5}{\Gamma(K_L \to \pi^+\pi^- e^+ e^-)}~.
\end{equation} 
In terms of the invariant form factors defined in (\ref{ff}), the
asymmetry is given by \cite{PT68,BCEG95} 
\begin{eqnarray} 
{\cal A}_{CP} &=& \displaystyle\frac{e^2}{128 \pi^8 M_{K^0}
\Gamma(K_L \to \pi^+\pi^- e^+ e^-)} \label{asymm} \\
& & \int d_{LIPS} \ve_{\mu\nu\rho\sigma}p_1^\mu p_2^\nu k_+^\rho k_-^\sigma
\frac{(p_1-p_2)\cdot(k_+-k_-)}{q^4}{\rm sgn}(\sin\Phi\cos\Phi)
\IM[(E_1-E_2)M^*] \no
\end{eqnarray} 
where $d_{LIPS}$ is the invariant phase space integration measure.
We will also consider the asymmetry for certain cuts (in
$q^2$). In this case, both the phase space integration in the numerator
of (\ref{asymm}) and $\Gamma(K_L \to \pi^+\pi^- e^+ e^-)$ in the 
denominator must be modified accordingly.

When CP is conserved we have $E_2(p_1,p_2,q)=E_1(p_2,p_1,q)$.
Therefore, only CP-violating contributions to $E_1,
E_2$ matter for the numerator in (\ref{asymm}).

The most recent published result for the asymmetry comes
from the KTeV-Collaboration \cite{ktev00}:
\begin{equation} 
{\cal A}_{CP} = 13.6 \pm 2.5 ~({\rm stat}) \pm 1.2 ~({\rm syst}) ~\%~.
\label{exp:ktev}
\end{equation} 
The preliminary result from NA48 \cite{na4800} is fully
compatible with this value.

Before delving into the theoretical analysis, we comment on the
observed sign of the asymmetry (\ref{exp:ktev}). With our conventions
\cite{BCEG95} and with the assumption that
$\IM[(E_1-E_2)M^*]$ has a unique sign all over phase space, the
theoretical expression (\ref{asymm}) implies 
\begin{equation} 
{\rm sgn}~{\cal A}_{CP} = {\rm sgn}~\IM[(E_1-E_2)M^*]~.
\label{sign}
\end{equation}
As we shall show in the sequel, $\IM[(E_1-E_2)M^*]$ does in fact have
a unique sign and must therefore be positive according to
Eqs.~(\ref{exp:ktev}) and (\ref{sign}).

\paragraph{3.} 
At lowest order in the low-energy expansion, the electric amplitudes
correspond to pure Bremsstrahlung:
\begin{eqnarray} 
E_1 &=& \displaystyle\frac{-2ie A^{\rm tree}(K_L\to \pi^+\pi^-)}{q^2+2
p_1\cdot q} = \displaystyle\frac{-2\sqrt{2}e \eta_{+-}A_0^{\rm tree}}
{q^2+2 p_1\cdot q} \nn
E_2 &=&- E_1(p_1 \to p_2)~.
\label{Ep2}
\end{eqnarray} 
From the dominant $\Delta I=1/2$ weak Hamiltonian,
the $I=0$ amplitude $A_0^{\rm tree}$ for $K_S \to \pi\pi$ 
decays is given by
\begin{equation} 
A_0^{\rm tree} = \sqrt{2}G_8 F (M_K^2-M_\pi^2)~, 
\end{equation} 
with $G_8=9.1\cdot 10^{-6}$ GeV$^{-2}$ and $F=F_\pi=92.4$ MeV at 
tree level. The current values for the CP-violating quantity 
$\eta_{+-}$ are \cite{pdg00}
\begin{eqnarray} 
|\eta_{+-}| &=& (2.276 \pm 0.017)\cdot 10^{-3} \nn
\Phi_{+-} &=& {\rm arg}~\eta_{+-} = (43.3 \pm 0.5)^\circ ~.
\label{etaerr}
\end{eqnarray}

The magnetic tree-level amplitude starts at $O(p^4)$ where it is 
completely given in terms of low-energy constants of the 
nonleptonic weak Lagrangian of $O(p^4)$ \cite{KMW90,EKW93}. In the 
notation of Ref.~\cite{BEP92}, the leading-order magnetic amplitude 
is written as
\begin{equation} 
M = - \frac{eG_8}{2\pi^2 F}(a_2 + 2a_4)
\label{Mtree}
\end{equation}
with dimensionless coefficients $a_2,a_4$ of order one. As shown in 
Ref.~\cite{BEP92}, the
chiral anomaly induces positive contributions to these coefficients.
If the anomaly-induced contributions were the dominant ones,
we would expect $a_2 + 2a_4$ to be positive.

It is easy to check for the tree-level amplitude that the
CP-violating quantity $\IM[(E_1-E_2)M^*]$ has a definite sign
that equals the sign of $a_2 + 2a_4$. From
(\ref{exp:ktev}) and (\ref{sign}) we would therefore conclude that 
$a_2 + 2a_4$ is positive supporting the hypothesis that the
contributions from the chiral anomaly dominate the 
magnetic amplitude (\ref{Mtree}).

However, this cannot be the whole story. Assuming a constant magnetic
amplitude, one can extract this amplitude from the branching ratio for 
$K_L \to \pi^+ \pi^- \gamma$. Following Sehgal and Wanninger
\cite{SW92} (the same procedure is used in Refs.~\cite{HS93,ESW95}),
one obtains in their notation
\begin{equation} 
M = - \frac{0.76 e |f_s|}{M_K^4} ~,
\end{equation} 
with $|f_s|=3.9\cdot 10^{-7}$ GeV denoting the absolute value of the 
$K_S\to \pi^+\pi^-$
amplitude. The resulting asymmetry at tree level is
\begin{equation} 
{\cal A}_{CP}^{\rm tree} = 7.7~\% ~,
\end{equation} 
much smaller than the experimental result (\ref{exp:ktev}). 
Clearly, higher-order corrections are necessary to
understand the asymmetry.

\paragraph{4.} 
Chiral corrections to the tree-level amplitudes were already
considered in 
Ref.~\cite{ESWW96}. Those authors included one-loop corrections to the
absorptive parts of both electric and magnetic form factors. Whereas
the corrections are small for the magnetic amplitude they are sizable
for the electric form factors, mainly due to the large final state
interactions of two pions with $I=0$ in an $S$-wave, as first
pointed out by Sehgal and Wanninger \cite{SW92}. To some extent,
Elwood et al. \cite{ESWW96} compensated the neglect of dispersive 
contributions by using the tree-level value of the weak coupling
constant $G_8$ also in the loop amplitude. On the other hand, this 
is partly double counting because $G_8$ should be reduced by about 
30 $\%$ \cite{KMW91,KDHMW92} when effects of $O(p^4)$ are included 
in the amplitude for $K_1^0 \to \pi^+\pi^-$.

We propose here a different procedure for the CP-violating electric 
amplitude. We first decompose the electric form factors into a 
Bremsstrahlung part and a direct-emission piece:
\begin{equation} 
E_i = E_i^{\rm B} + E_i^{\rm DE} \qquad (i=1,2)~.
\end{equation} 
In the first part, we use the phenomenological $K_S \to \pi^+ \pi^-$
amplitude,
\begin{eqnarray} 
E_1^{\rm B} &=& - \displaystyle\frac{2\sqrt{2}e \eta_{+-}}{q^2+2p_1\cdot q}
\left[A_0 e^{i\delta_0^0} +\frac{1}{\sqrt{2}}A_2 e^{i\delta_0^2}\right]
\nn
E_2^{\rm B} &=& - E_1^{\rm B}(p_1 \to p_2),\label{EiB}
\end{eqnarray} 
with isospin amplitudes $A_0=2.71\cdot 10^{-7}$ GeV, 
$A_2=0.12\cdot 10^{-7}$ GeV taken
from experiment. The pion-pion phase shifts $\delta_0^0, \delta_0^2$ at 
$s=M_K^2$ are taken from a new analysis of pion-pion scattering combining
dispersion theory (Roy equations) with chiral perturbation theory  
\cite{ACGL00,CGL00}:
\begin{equation} 
\delta_0^0 = 39.1^\circ ~, \qquad \delta_0^2 = - 8.5^\circ ~.
\end{equation}

For the CP-violating direct-emission part, we have calculated the 
amplitude of
$O(p^4)$ for $K_1^0 \to \pi^+ \pi^- \gamma^*$, including both a
local amplitude derived from the nonleptonic weak Lagrangian of
$O(p^4)$ \cite{KMW90,EKW93} and a loop amplitude restricted to the
dominant pion loops. Of course, one has to project
out the Bremsstrahlung part of $O(p^4)$ that is already included in 
$E_i^{\rm B}$ in (\ref{EiB}). Altogether, we have
\begin{eqnarray} 
E_1^{\rm DE} &=& E_1^{\pi-{\rm loops}}(\mu=M_\rho) \nn
& & + \frac{2 e \eta_{+-} G_8}{3F}q^2\left[N^r_{14}(M_\rho)-N^r_{15}(M_\rho)-
3(N^r_{16}(M_\rho)+N_{17})\right]\nn
&&+\frac{4 e \eta_{+-} G_8}{F}p_2\cdot q
\left[N^r_{14}(M_\rho)-N^r_{15}(M_\rho)-N^r_{16}(M_\rho)-N_{17}\right]
\nn
E_2^{\rm DE} &=& - E_1^{\rm DE}(p_1\leftrightarrow p_2) ~.\label{EiDE}
\end{eqnarray} 
The explicit form of the direct-emission pion-loop amplitude 
$E_1^{\pi-{\rm loops}}(\mu=M_\rho)$ 
can be found in Ref.~\cite{Diss}. Because we have only included the
dominant pion loops there is a residual, but numerically
unimportant scale dependence in (\ref{EiDE}). We have chosen the
usual renormalization scale $\mu=M_\rho$. 
The numerical values for the low-energy constants $N^r_i(M_\rho)$
($N_{17}$ is scale independent) are taken from a recent analysis 
\cite{Diss,Pichl00} of the branching ratio $B(K_L \to \pi^+\pi^- e^+ e^-)$. 
Anticipating the numerical discussion, the direct-emission form
factors $E_i^{\rm DE}$ are negligible for ${\cal A}_{CP}$, especially 
when integrated over the whole phase space. 

For the magnetic amplitude, we must be less ambitious for the time
being. Already
at $O(p^4)$, we cannot claim to be able to calculate the coefficients
in Eq.~(\ref{Mtree}). Moreover, $\eta-\eta^\prime$ mixing is known to
produce a big contribution of $O(p^6)$ \cite{ENP94}
that interferes destructively with the amplitude induced by the chiral
anomaly. In addition, there are two apparently non-equivalent versions 
of implementing vector and axial-vector exchange in nonleptonic weak 
transitions \cite{EPR90,DP97}. 

On the other hand, there exists strong experimental evidence for
higher-order effects parametrized in terms of a $\rho$-dominated form
factor \cite{ktev00,Ramberg}. Including final state interactions
appropriate for $P$-wave $\pi\pi$ scattering, we therefore adopt the
magnetic amplitude measured by KTeV,
\begin{equation} 
M = \displaystyle\frac{e |f_S|}{M_K^4}\tilde g_{M1}
e^{i\delta_1^1(s_\pi)} \left[1 + \displaystyle\frac{a_1/a_2}
{M_\rho^2-M_K^2+2 M_K E_\gamma^*}\right] \label{Mexp}~,
\end{equation}  
with $E_\gamma^*$ the total lepton energy in the kaon rest frame and
with\footnote{The quantity $a_2$ in (\ref{Mexp}) must not be confused
with the same expression in (\ref{Mtree}).}  \cite{ktev00}
\begin{equation} 
\tilde g_{M1} = 1.35 \begin{array}{l} + 0.20 \\ - 0.17 \end{array} ~,
\qquad a_1/a_2 = (- 0.720 \pm 0.028) ~{\rm GeV}^2 ~. \label{Mpar}
\end{equation} 
The $I=J=1$ phase shift is parametrized as
\begin{equation} 
\delta_1^1(s_\pi) = \frac{1}{2}\arcsin \left[\frac{4 q_\pi^3}
{\sqrt{s_\pi}}(a_1^1 + b_1^1 q_\pi^2 + c_1^1 q_\pi^4)\right]
\end{equation} 
with
\begin{equation} 
\begin{array}{lll}
a_1^1 = 0.038/M_{\pi^+}^2~, & b_1^1 = 0.0057/M_{\pi^+}^4~, &
c_1^1 = 0.001/M_{\pi^+}^6 ~,\\ 
& q_\pi=(s_\pi/4 - M_{\pi^+}^2)^{1/2}~. &
\end{array}   
\end{equation} 
The slope parameters $a_1^1,b_1^1$ are taken from
Refs.~\cite{ACGL00,CGL00} and the coefficient $c_1^1$ is introduced to
reproduce the correct phase shift at $s=M_K^2$.
  
With the sign convention of (\ref{EiB}), the magnetic coupling 
$\tilde g_{M1}$ must be positive in order to
reproduce the measured sign of the asymmetry.
Unfortunately, there is at present no unique way to
infer from (\ref{Mexp}) and the experimental values (\ref{Mpar})
the sign of the lowest-order combination $a_2 + 2a_4$ in
(\ref{Mtree}). A recent analysis of $K\to \pi\pi\gamma$ transitions by
D'Ambrosio and Gao \cite{DG00} in the framework of the vector-field
representation for (axial-)vector resonances finds that both signs are
possible depending on the coupling strength of spin-1 exchange. A
deeper understanding of higher-order effects in nonleptonic
weak interactions will be necessary before one could claim that the
chiral anomaly is largely responsible for the measured sign of the 
asymmetry.

\paragraph{5.}  
With electric form factors given in Eqs.~(\ref{EiB}) and (\ref{EiDE})
and with the magnetic amplitude (\ref{Mexp}) we obtain for the total
integrated asymmetry
\begin{equation} 
{\cal A}_{CP} = 13.7~\% 
\label{ACPtheor}
\end{equation} 
if all input quantities are taken at their respective mean values.

This value agrees with previous theoretical estimates and
with the experimental results from KTeV and NA48. The main issue
we want to address here is the theoretical uncertainty of this
prediction. Electric and magnetic amplitudes are on quite a
different footing in this respect. Except for the $P$-wave phase shift 
$\delta_1^1(s_\pi)$ that we will lump together with the phases
occurring in $E_i^B$, we cannot ascribe a meaningful theoretical error 
to the magnetic amplitude. We will therefore scale the prediction
(\ref{ACPtheor}) to the measured magnetic coupling $\tilde g_{M1}$
and include the experimental error of the ratio $a_1/a_2$
explicitly. Accounting for the measured branching ratio in the same
way, we arrive at 
\begin{equation} 
{\cal A}_{CP} = \displaystyle\frac{3.63\cdot 10^{-7}}
{B(K_L \to \pi^+\pi^- e^+ e^-)}\cdot \displaystyle\frac
{\tilde g_{M1}}{1.35}\left(13.7 \pm 1.3 \right)~\% 
\label{ACPerr1}
\end{equation} 
where the given error is due to the error of $a_1/a_2$ in 
(\ref{Mpar}) only. Future experimental improvements and
correlations between the measured values for 
$B(K_L \to \pi^+\pi^- e^+ e^-)$, $\tilde g_{M1}$ and $a_1/a_2$ can
easily be incorporated in this formula.

In contrast, the uncertainty of ${\cal A}_{CP}$ related to the electric
form factors is fully under control theoretically. For the total asymmetry 
under consideration here, the CP-violating
direct-emission amplitude (\ref{EiDE}) is completely negligible and
does not affect ${\cal A}_{CP}$ to the accuracy given. Of course, this 
is a consequence of the dominance of the Bremsstrahlung form 
factors (\ref{EiB}) at small $q^2$. 

Inclusion of the $I=2$ amplitude $A_2$ is itself a small effect
contributing 0.25 $\%$ to ${\cal A}_{CP}$. The relatively big errors of
$A_2$ and $\delta_0^2$ due to isospin violation \cite{mumd} and
electromagnetic corrections \cite{CDG00} have therefore no impact on
${\cal A}_{CP}$. On the other hand, those corrections are small for 
$A_0$ and $\delta_0^0$ that do matter for the asymmetry. Assigning an
uncertainty $\pm 0.05 \cdot 10^{-7}$ GeV to $A_0$ induces 
$\Delta {\cal A}_{CP}=0.25 ~\%$. Likewise, the error for $|\eta_{+-}|$
given in (\ref{etaerr}) propagates into $\Delta {\cal A}_{CP}=0.10 ~\%$. 

Because of the dominance of Bremsstrahlung, the phases relevant for 
${\cal A}_{CP}$ enter effectively in the combination $\delta_0^0
- \delta_1^1(s_\pi) + \Phi_{+-}$. From the recent analysis of pion-pion
phase shifts \cite{ACGL00,CGL00} we extract an uncertainty of
$0.6^\circ$ for $\delta_0^0 - \delta_1^1(s_\pi)$. Together with the
error of $\Phi_{+-}$ displayed in (\ref{etaerr}), the conservative uncertainty 
$\Delta\left(\delta_0^0 - \delta_1^1(s_\pi) + \Phi_{+-}\right)=1^\circ$
gives rise to $\Delta {\cal A}_{CP}=0.05 ~\%$. 

Combining the uncertainties from $A_0$, $|\eta_{+-}|$ and from the phases in
quadrature, we add the resulting theoretical error of $0.3~\%$ to the
previous one due to $a_1/a_2$ to obtain our final result for the total 
asymmetry:
\begin{equation} 
{\cal A}_{CP} = \displaystyle\frac{3.63\cdot 10^{-7}}
{B(K_L \to \pi^+\pi^- e^+ e^-)}\cdot \displaystyle\frac
{\tilde g_{M1}}{1.35}\left(13.7 \pm 1.3 \pm 0.3\right)~\% ~.
\label{ACPfinal}
\end{equation} 
Given the magnetic amplitude and the branching ratio $B(K_L \to 
\pi^+\pi^- e^+ e^-)$, the CP-violating asymmetry can be calculated
to a relative accuracy of about $2~\%$.

\paragraph{6.}  
Following the authors of Refs.~\cite{ESW95,ESWW96,Sav99}, we consider 
the asymmetry also for different cuts in $q^2$, the invariant mass squared of
the leptons. For this purpose, the chiral amplitude to $O(p^4)$ 
\cite{ESW95,Pichl00} is used to calculate the rate 
$\Gamma(K_L \to \pi^+\pi^- e^+ e^-)$ in the denominator of the 
asymmetry for the cuts in question.
As already mentioned, the low-energy constants $N_i^r$ are taken from 
a recent analysis \cite{Diss,Pichl00} of the total branching ratio 
$B(K_L \to \pi^+\pi^- e^+ e^-)$.

Suppressing errors, the results are collected in Table
\ref{tab:cuts}. One immediate conclusion is that the asymmetry is 
maximal for $q^2 \grts (10~{\rm MeV})^2$.

\begin{table}[ht]
\begin{center}
\caption{The asymmetry ${\cal A}_{CP}$ for different
cuts in $q^2$, the invariant mass squared of the lepton pair.}
\label{tab:cuts}
\vspace{.5cm}
\begin{tabular}{|c||c|}\hline
$q^2 > [\mbox{MeV}^2]$ & ${\cal A}_{CP}$ in \% \\ 
\hline \hline
entire phase space & 13.7 \\ \hline \hline
$2^2$ & 15.2\\ \hline
$10^2$ & 15.7\\ \hline
$20^2$ & 14.0\\ \hline
$30^2$ & 12.1\\ \hline
$40^2$ & 10.2 \\ \hline
$60^2$ & 7.2\\ \hline
$80^2$ & 4.8 \\ \hline
$100^2$ & 3.2\\ \hline
$120^2$ & 2.0\\ \hline
$180^2$ & 0.4\\ \hline \hline
\end{tabular}
\end{center}
\end{table}

Compared to the previous situation for the total asymmetry with
$q^2 \ge 4 m_e^2$, there are now additional uncertainties that 
increase with the lower limit on $q^2$. In the numerator of the 
asymmetry, the direct-emission form factors $E_i^{\rm DE}$ 
in (\ref{EiDE}) become 
more important in comparison with the Bremsstrahlung amplitude. 
Likewise, the rate in the denominator becomes more sensitive to 
the inaccurately known weak low-energy constants $N_i^r$ as one 
moves to larger $q^2$.

For large $q^2$, both the asymmetry  and the rate decrease rapidly. 
For illustration, we therefore choose a realistic cut of 
$q^2 > (40~{\rm MeV})^2$. The CP-conserving amplitude of $O(p^4)$
depends on the combination $Z(\mu)=N_{14}^r(\mu)-N_{15}^r(\mu)-
3(N_{16}^r(\mu)-N_{17})$ of weak low-energy constants. With the 
mean value $Z(M_\rho)=0.023$ extracted from the total 
branching ratio \cite{Pichl00}, about 13~$\%$ of all events 
(almost 16~$\%$ of those with $q^2 > (2~{\rm MeV})^2$ as in the 
KTeV experiment \cite{ktev00}) satisfy $q^2 > (40~{\rm MeV})^2$. 

\renewcommand{\arraystretch}{1.5}
\begin{table}[ht]
\begin{center}
\caption{Dependence of the rate $\Gamma_{\rm cut}
(K_L \to \pi^+\pi^- e^+ e^-)$ on the electric amplitude
\protect\cite{Pichl00} for $q^2>(40~{\rm MeV})^2$. 
$E_i^{O(p^4)}$ denotes the complete electric amplitude to
$O(p^4)$ that depends on the combination $Z(\mu)=N_{14}^r(\mu)-
N_{15}^r(\mu)- 3(N_{16}^r(\mu)-N_{17})$ of weak low-energy constants.
The magnetic amplitude is given in (\protect\ref{Mexp}).}
\label{tab:p4}
\vspace{.5cm}
\begin{tabular}{|c||c|}\hline
 & $\Gamma_{\rm cut}/\Gamma_{\rm cut}(O(p^4))$ \\ 
\hline \hline
$E_i^B$ & 0.60 \\ \hline
$E_i^{O(p^4)}$ with $Z(M_\rho)=0.023$ & 1. \\ \hline
$E_i^{O(p^4)}$ with $Z(M_\rho)=0$ &  0.75 \\ 
\hline \hline
\end{tabular}
\end{center}
\end{table}

However, these numbers are rather sensitive to the precise value
of $Z(M_\rho)$. In Table \ref{tab:p4}, the dependence
of the rate on the electric amplitude is exhibited for $q^2 > (40~{\rm
MeV})^2$. About $25\%$ of the rate is due to $Z(M_\rho)$ 
being different from zero. This also documents that 
$Z(M_\rho)$ can  in principle be extracted with much
higher precision from the partial rate than from the total branching
ratio. In the sample accumulated by KTeV \cite{Cox}, several hundred
events are expected to satisfy $q^2 > (40~{\rm MeV})^2$.

As already mentioned, the direct-emission amplitude of $O(p^4)$ also
enters the numerator of the asymmetry. In contrast to the rate where
the CP-conserving electric amplitude of $O(p^4)$ overtakes the 
CP-violating Bremsstrahlung amplitude (\ref{Ep2}) of $O(p^2)$
already at fairly small $q^2$ \cite{Pichl00}, only the CP-violating 
part of direct emission matters 
in the numerator. For the considered cut $q^2 > (40~{\rm MeV})^2$ this 
is still only a small correction to the leading-order Bremsstrahlung. 
Numerically, the correction is smaller than the uncertainty due 
to $A_0$ and it will be included in the theoretical error.

Consequently, for reasonable cuts in $q^2$ the asymmetry can still be 
predicted rather precisely in terms of the magnetic amplitude and the 
branching ratio. For $q^2 > (40~{\rm MeV})^2$, we obtain 
\begin{equation} 
{\cal A}_{CP} = \displaystyle\frac{4.75\cdot 10^{-8}}
{B(K_L \to \pi^+\pi^- e^+ e^-; q^2>(40~{\rm MeV})^2)}
\cdot \displaystyle\frac
{\tilde g_{M1}}{1.35}\left(10.2 \pm 1.1 \pm 0.3\right)~\% 
\label{ACPcut}
\end{equation} 
where the first error is again due to the error of $a_1/a_2$ in 
(\ref{Mpar}), the second one being the genuine theoretical error.

\paragraph{7.}  
The electric direct-emission amplitude for the decay 
$K_L \to \pi^+\pi^- e^+ e^-$ has been calculated in chiral 
perturbation theory. The CP-conserving part is important for the rate
because it dominates the CP-violating Bremsstrahlung amplitude already
for rather low invariant masses of the lepton pair. With appropriate
cuts in $q^2$, it will be possible to extract the relevant combination 
of low-energy constants in the CP-conserving amplitude of $O(p^4)$
with good precision.

In contrast, the CP-violating asymmetry is quite insensitive to 
electric direct emission. This allows for an accurate
calculation of the asymmetry once the magnetic amplitude and branching
ratio have been determined experimentally. For realistic cuts in $q^2$,
the asymmetry can be predicted with a relative accuracy of 
$2 \div 3 \%$.

The contribution of the chiral anomaly to the magnetic amplitude of 
$O(p^4)$ leads to the observed sign of the asymmetry. However,
higher-order terms in the magnetic amplitude, clearly required by
experiment but not reliably calculable in chiral perturbation theory,
make this connection less conclusive.

\vfill 
\noindent 
We thank T. Barker, B. Cox, G. D'Ambrosio, M. Jeitler, S. Ledovskoy
and L.M. Sehgal for helpful correspondence.

\end{document}